\documentclass[twocolumn]{aastex63}
\usepackage[version=4]{mhchem}
\usepackage{graphicx}
\usepackage{multirow}
\usepackage{bm}

\newcommand{\lamD}{$\lambda / D$ }

\received{2023 August 16}
\revised{2023 September 20}
\accepted{2023 September 25}
\submitjournal{AJ}

\shortauthors{Currie et al.}

\begin{document}

\title{Mitigating Worst-Case Exozodiacal Dust Structure in High-contrast Images of Earth-like Exoplanets}

\correspondingauthor{Miles Currie}
\email{mcurr@uw.edu}

\author[0000-0003-3429-4142]{Miles H. Currie}
\affiliation{Department of Astronomy and Astrobiology Program, University of Washington, Box 351580, Seattle, Washington 98195, USA}
\affiliation{NASA Nexus for Exoplanet System Science, Virtual Planetary Laboratory Team, Box 351580, University of Washington, Seattle, Washington 98195, USA}

\author{Christopher C. Stark}
\affiliation{NASA Goddard Space Flight Center, Greenbelt, MD 20771, USA}

\author[0000-0003-2769-0438]{Jens Kammerer}
\affiliation{Space Telescope Science Institute, 3700 San Martin Drive, Baltimore, MD 21218, USA}

\author{Roser Juanola-Parramon}
\affiliation{NASA Goddard Space Flight Center, Greenbelt, MD 20771, USA}

\author[0000-0002-1386-1710]{Victoria S. Meadows}
\affiliation{Department of Astronomy and Astrobiology Program, University of Washington, Box 351580, Seattle, Washington 98195, USA}
\affiliation{NASA Nexus for Exoplanet System Science, Virtual Planetary Laboratory Team, Box 351580, University of Washington, Seattle, Washington 98195, USA}
\affiliation{Astrobiology Center, 2-21-1 Osawa, Mitaka, Tokyo 181-8588, Japan}

\begin{abstract}
Detecting Earth-like exoplanets in direct images of nearby Sun-like systems brings a unique set of challenges that must be addressed in the early phases of designing a space-based direct imaging mission. In particular, these systems may contain exozodiacal dust, which is expected to be the dominant source of astrophysical noise. Previous work has shown that it may be feasible to subtract smooth, symmetric dust from observations; however, we do not expect exozodiacal dust to be perfectly smooth. Exozodiacal dust can be trapped into mean motion resonances with planetary bodies, producing large-scale structures that orbit in lock with the planet. This dust can obscure the planet, complicate noise estimation, or be mistaken for a planetary body. Our ability to subtract these structures from high-contrast images of Earth-like exoplanets is not well understood. In this work, we investigate exozodi mitigation for Earth--Sun-like systems with significant mean motion resonant disk structures. We find that applying a simple high-pass filter allows us to remove structured exozodi to the Poisson noise limit for systems with inclinations $< 60^\circ$ and up to 100 zodis. However, subtracting exozodiacal disk structures from edge-on systems may be challenging, except for cases with densities $<5$ zodis. For systems with three times the dust of the Solar System, which is the median of the best fit to survey data in the habitable zones of nearby Sun-like stars, this method shows promising results for mitigating exozodiacal dust in future HWO observations, even if the dust exhibits significant mean-motion resonance structure. 
\end{abstract}


\section{Introduction} \label{sec:intro}

Stars are not solitary objects; they host a complex system that may include a variety of planets, comets, asteroids, and a sea of small debris generated from larger bodies known as an exozodiacal disk. As we plan for a new era of detecting and characterizing Earth-like planets via high-contrast imaging, it is imperative to define the impact of all astrophysical sources that may contribute to the noise budget of future observations. In particular, exozodiacal dust may dominate the noise budget of a given system. Our ability to fit and subtract this dust from directly imaged systems containing Earth-like exoplanets will depend on properties of both the debris disk and the observatory. 

Our own Solar System (SS) provides a nearby and well-studied example of dust in the habitable zone, known as zodiacal dust. With a surface brightness of $\sim22$ mag/arcsec$^2$ at 1 AU in the V band \citep{Levine2006}, zodiacal dust is a non-negligible source of noise for all astronomical observations \citep[e.g.][]{Dermott2002}.  Other stars host dust systems known as exozodiacal dust that can be much brighter than zodiacal dust, introducing an additional source of noise when observing exoplanets  \citep{Roberge2012-te}.

The origin of exozodiacal dust (exozodi) for a typical system is not well understood. Exozodi may originate from distant objects, analogous to our Solar System's Kuiper Belt, whose dust slowly migrates inward to the habitable zone via Poynting-Robertson drag \citep{Reidemeister2011, Kennedy2015}.  This dust may also be generated by eccentric comets evaporating near periastron \citep{Beust1990}, a separate population of warm planetesimals similar to our Asteroid Belt, or a recent catastrophic event that redistributed material to the habitable zone \citep{Weinberger2011}. One or more of these processes may generate exozodiacal dust in Sun-like stellar systems, leading to variation in the observed density \citep{Ertel2020-kt}. Regardless of its origin, exozodiacal dust can obscure observations of Earth-like exoplanets, and will likely need to be subtracted from the images. 

Exozodiacal dust can populate the warm inner regions of planetary systems, where Earth-like planets may reside. A recent survey using the Large Binocular Telescope Interferometer suggests that the median level of habitable zone exozodiacal dust for nearby Sun-like stars is approximately three zodis \citep{Ertel2020-kt}, where one zodi is equal to the SS level of zodiacal dust in the habitable zone. Exozodi mitigation is therefore a particularly prudent consideration for precursor studies supporting a future Habitable Worlds Observatory \citep{Astro2020}, which will be designed to detect and characterize Earth-like exoplanets in the habitable zone via high-contrast direct imaging.

Exozodiacal dust can impact exoplanet detection, and removing exozodi from observations of Earth-like exoplanets may be challenging, especially if the disk is spatially inhomogeneous. Our ability to remove exozodi will depend primarily on the disk's brightness, the scale of the instrument's PSF, and the method used fit the exozodi's spatial distribution. If exozodi is smooth, it may be fairly straightforward to fit a high-order polynomial to the observed image to subtract off the bulk of the dust---this method is only effective up to a few tens of zodis, after which it is no longer possible to subtract the background down to the Poisson noise limit \citep{Kammerer2022-vh}. However, we do not expect exozodi to be perfectly smooth. The SS zodiacal cloud has features associated with specific asteroid families, and the Earth is known to shepherd dust into a clumpy, circumsolar resonant ring structure \citep[e.g.][]{Dermott1985, Dermott1994-af, Reach1995}. A similar ring structure has also been observed near Venus's orbit \citep{Stenborg2021}. Furthermore, the outer regions of debris disks observed around other stars exhibit clumps, warps, rings, and gaps \citep[e.g.][]{Greaves1998, Wilner2002, Kalas2005}. Analogous structures may exist in the inner regions of disks; one possible morphology is an annulus around the star at the orbital radius of the planet, with a width of a few tenths of an AU and a gap at the location of the planet \citep{Kuchner2003}. These structures may be difficult to remove from observations, preventing us from detecting potentially habitable planets \citep{Roberge2012-te}. Although preliminary studies have suggested exozodi may significantly impact planetary detection \citep[e.g.][]{Defrere2012}, the feasibility of removing these structures from high-contrast images has not been thoroughly investigated.

To date, most exoplanet yield studies assume that we are able to subtract exozodiacal dust down to the Poisson noise limit \citep[e.g.][]{Brown2005, Stark2014-bj, Savransky2016, Gaudi2020, LUVOIR2019}. While this appears roughly valid for smooth disks with densities $< 30$ zodis \citep{Kammerer2022-vh}, we do not know if this is the case for disks with structures. In this work, we simulate observations of planetary systems with exozodiacal disk structures and test our ability to subtract down to the Poisson noise limit using a high-pass filtering technique. We consider systems covering a range of inclinations and zodi levels up to 100 times the SS zodiacal dust level, and test our ability to detect an exoplanet in the post-processed images.
To help inform trades for the required mirror diameter for the nominally $\sim$6 m inscribed diameter Habitable Worlds Observatory---the top recommendation for the flagship mission of the Astro2020 Decadal Survey \citep{Astro2020}---we examine two possibilities for primary mirror size. We consider an 8 m circumscribed diameter mirror with an inscribed diameter similar to the Decadal recommendation, as well as a larger 12 m option.

In Section~\ref{sec:methods}, we present our methods for generating astrophysical scenes, synthesizing coronagraph observations, subtracting the disk structure, and estimating the resulting signal-to-noise ratio of an injected Earth-like exoplanet. In Section~\ref{sec:results}, we present our results for a grid of simulations. In Section~\ref{sec:discussion}, we discuss our results and the lessons learned, then conclude in Section~\ref{sec:conclusion}.

\section{Methods}\label{sec:methods}
To investigate how exozodiacal disk structure affects our ability to extract planetary signal, we adopt worst-case scenario models of gravitational mean motion resonant rings created by Earth twins in exozodiacal disks, simulate images of the coronagraph response including stellar speckles, and add photon noise to the simulated observations. We then process the images by applying a high-pass filter to remove residual exozodiacal structure from PSF-subtracted images, and apply methods to detect the planetary signal. We quantify the performance of our technique by analyzing the residual noise in the post-processed image.

\subsection{Simulating debris disk images}\label{sec:disk_sims}

\subsubsection{N-body Models}
\begin{figure*}
    \centering
    \includegraphics{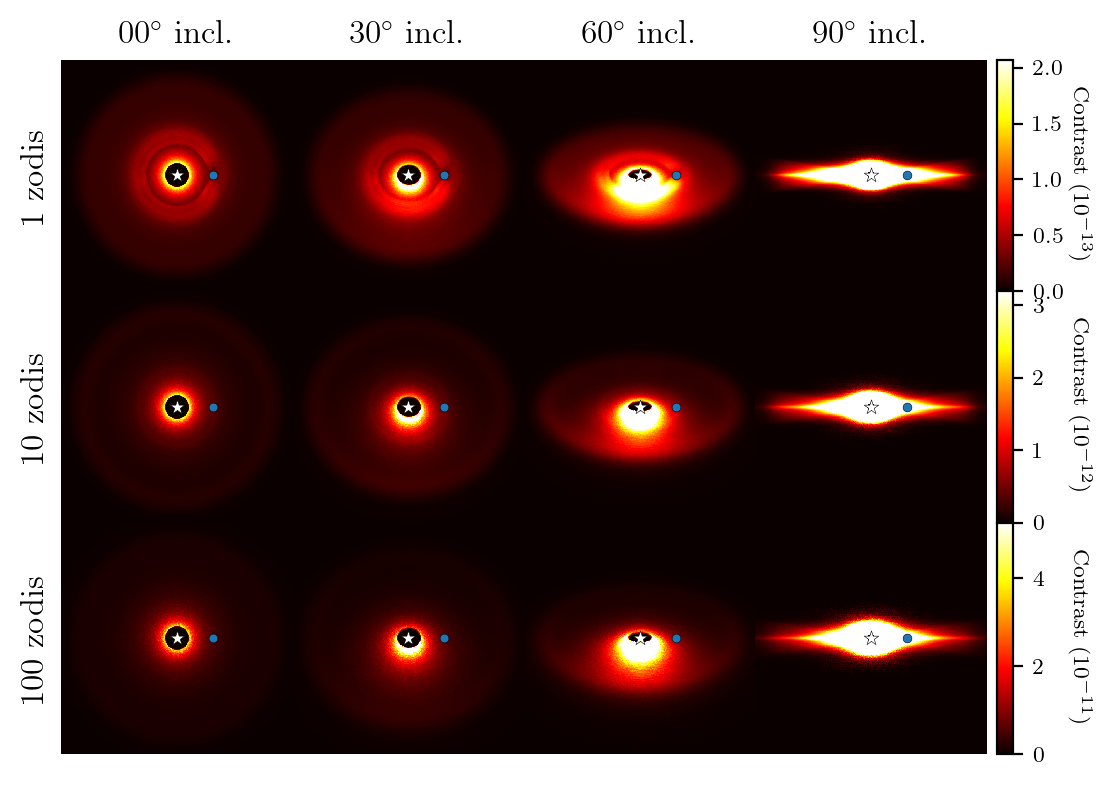}
    \caption{A sample of scattered light images of exozodiacal dust disks including worst-case-scenario resonant structure used in this work. Each row represents a disk with 1, 10, or 100 zodis and includes a colorbar representing the contrast of 0, 30, 60, and 90 degree inclined disks. The blue dot represents an Earth-like planet at 1 AU away from the star, and the star is located in the center of the image. The structure appears less pronounced for the $>1$ zodi cases because it accounts for a smaller percentage of the total surface brightness due to enhanced collisional destruction of grains in denser disks. The disks in this figure are shown after the dithering and Mie theory artifact mitigation steps.}
    \label{fig:dustmap_ims}
\end{figure*}

We adopted the exozodiacal disk models of \citet{Stark2011-sv}, who simulate mean motion resonant ring structures created by planets around Sun-like stars for disks ranging from 1 to 100 zodis in density. These debris disk models were generated via n-body simulations, taking into account three-body gravitational dynamics between the star, a single planet, and a large population of dust grains, Poynting-Robertson and corpuscular drag, radiation pressure, and destructive collisions between dust grains. The models assumed a Dohnanyi size distribution at the moment of launch of the dust grains and self-consistently calculated the size distribution at all later points in time via collisional equilibrium \citep{Stark2009}. Notably, these models were specifically generated to represent a ``worst case scenario'' for mean motion resonant disk structures by tuning all of the physics to produce as much structure as possible. Specifically, these systems are composed of single planets on circular orbits around Sun-like stars and the parent bodies that generate the dust were placed at 2.5 to 3.0 times the semi-major axis of the planet to ensure as much dust as possible was delivered to the planet's mean motion resonant orbits via drag forces. Upon delivery, a fraction of the dust is gravitationally trapped in mean motion resonances, producing large-scale overdensities in the disk that orbit in lock with the planet. \citet{Kuchner2003} found these single-planet circular orbit scenarios typically exhibit asymmetries in the form of a density deficit, or ``gap'', at the location of the planet, and density enhancements or ``clumps'' both leading and trailing the planet, the former typically being slightly less dense.  From the library of models generated by \citet{Stark2011-sv}, we included those with Earth-mass planets at 1 AU and models with zodi levels of 1, 5, 10, 20, 50, and 100 zodis. Figure~\ref{fig:dustmap_ims} shows a sample of the debris disks used in this work, plotted after the dithering and Mie theory mitigation steps described Sections \ref{sec:dithering} and~\ref{sec:mie}, respectively.

\subsubsection{Generating images with \texttt{dustmap}}
\texttt{dustmap} \citep{Stark2011-sv} is an IDL suite designed to simulate density histograms, optical depth maps, thermal emission images, and scattered light images given a list of 3D particle locations. Each particle is assumed to represent a large number of dust grains, and we adopt the optical constants for astronomical silicates \citep{Draine1984} and use Mie theory to calculate the scattering efficiency and phase functions. In this work, we use \texttt{dustmap} to calculate scattered light images of these models for inclinations of $0^\circ$ (face-on), $30^\circ$, $60^\circ$, and $90^\circ$ (edge-on) with respect to the observer. We define the pixel scale to be $1.074$ mas at 500 nm, which allows us to bin the model pixels by integer values (avoiding interpolation) to achieve the resolution of our coronagraph models for both the 8 and 12 m telescope configurations. For the scattered light images, we assume that the disk is illuminated by a Sun-like star with stellar properties of 1 R$_\odot$, 1 L$_\odot$, T$_\mathrm{surface}$ $= 5770$ K, and log(g) $= 4.5$.

\subsubsection{Reducing particle noise}\label{sec:dithering}
\begin{figure}
    \centering
    \includegraphics[width=0.5\textwidth]{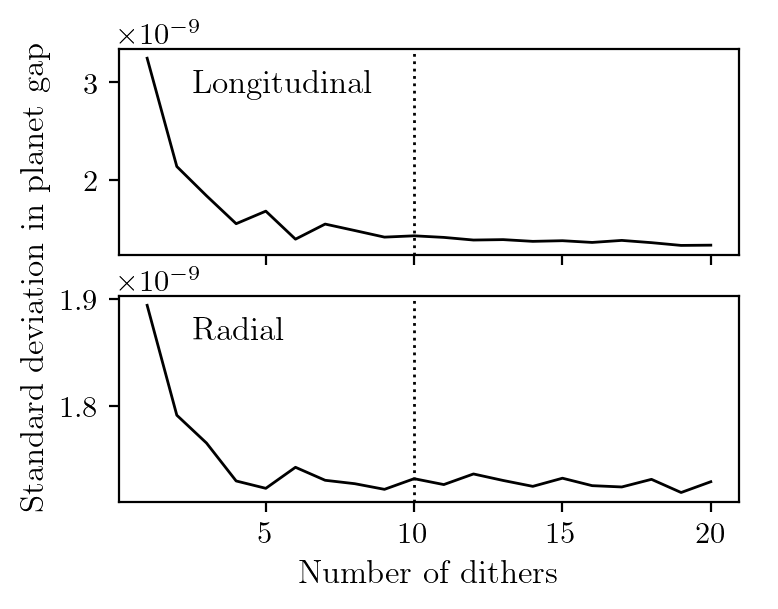}
    \caption{Standard deviation of a 7x7 pixel region centered on the planet location in the smoothed \texttt{dustmap} image as a function of the number of dithers in both the longitudinal (top panel) and radial (bottom panel) directions. For both dimensions, ten dithers is adequate to smooth over N-body particle noise in scattered light images generated using \texttt{dustmap}.}
    \label{fig:dither}
\end{figure}

Because N-body simulations are composed of a finite number of particles, the resulting \texttt{dustmap} outputs are not smoothly varying functions. This limitation in resolution introduces particle noise to the final simulation. To mitigate particle noise, we dither the image in both the longitudinal and radial directions, creating a series of images that vary slightly in longitude and magnification, and take the median of this series of images as our smoothed image. Dithering the disk in this fashion differs from the coronagraphic PSF convolution discussed later in Section~\ref{sec:coro} because it is a physical dither applied relative to the disk plane, which allows us to average over particle noise on sub-pixel scales. We find that ten dithers in each of the radial and longitudinal directions are required to adequately smooth the image, for a total of 100 dithers per exozodiacal disk model (see Figure~\ref{fig:dither}).

In the longitudinal direction, dithering is achieved by adjusting the longitude of the system in the \texttt{dustmap} call, effectively rotating the disk around the axis normal to the disk midplane. When generating scattered light maps, we run \texttt{dustmap} for an array of longitudes centered on the true longitude of the system spanning $5^\circ$. At the planet location, this $5^\circ$ span translates to a width slightly less than the PSF of the telescope.  The top panel of Figure~\ref{fig:dither} shows the standard deviation of a 7x7 pixel region centered at the location of where the planet would be in the disk simulation as a function of the number of dithers. We find that ten dithers in the longitudinal direction is adequate to stabilize the standard deviation of the region.

In the radial direction, we dither by adjusting the distance to the system in the \texttt{dustmap} call. Similar to longitudinal dithering, we run \texttt{dustmap} for an array of distances to the system centered on 10 pc, and spanning 0.02 pc. In our case, this span is sufficient to shrink or enlarge the scale of the image by one pixel, which translates to a fraction of the size of the PSF. Again, we find that ten dithers in the radial direction is adequate to stabilize the standard deviation of the region defined in the text above (see bottom panel of Figure~\ref{fig:dither}).

\subsubsection{Reducing Mie theory artifacts}\label{sec:mie}
\begin{figure}
    \centering
    \includegraphics[width=0.5\textwidth]{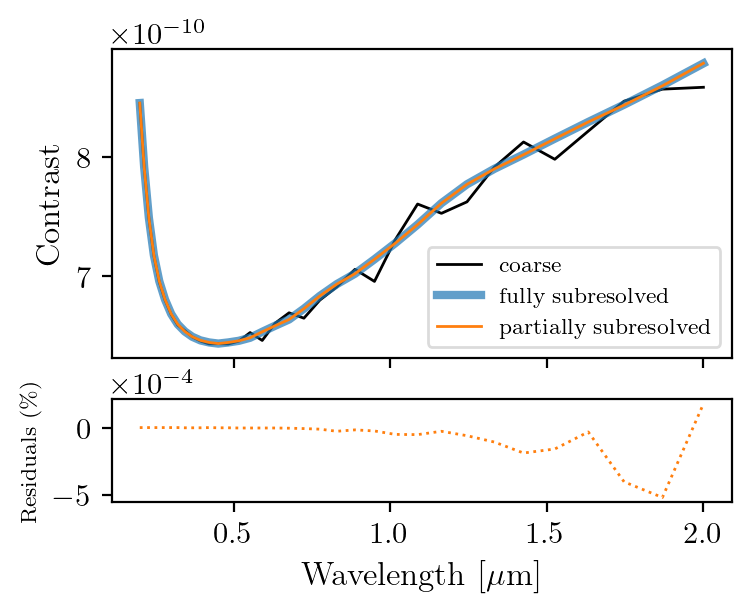}
    \caption{Upper panel: Contrast curves for different grain size lists. Ringing Mie theory artifacts are present when using the original coarse grain size list, and are reduced by using a subresolved or partially-subresolved grain size list. The run time of an individual dustmap call is significantly improved by using a partially subresolved grain size list. Lower panel: Fractional difference of the contrast curves produced by using the partially-subresolved grain size list and the fully-subresolved grain size list. The resulting contrast curve using the partially-subresolved grain size list exhibits negligible differences ($< 0.1 \%$) when compared to the contrast curve of the fully-subresolved grain size list.}
    \label{fig:particlesize}
\end{figure}

Mie theory assumes perfectly spherical grains. As a result, the calculated scattering coefficients and phase functions of a single grain size can feature unrealistic ``ringing,'' a well-known limitation of Mie theory. The original N-body simulations of \citet{Stark2011-sv} use a relatively coarse set of grain sizes for the dust in the system, with 25 grain sizes spanning 0.69 and 480 $\mu m$. Such a coarse grid does not sufficiently remove these ringing artifacts, which appear as visible discontinuities in the disk. These discontinuities in the contrast curve (see Figure~\ref{fig:particlesize}) would limit future studies of the impact of these exozodi models on spectral extraction, thus we opt to remove them. One option to remove these artifacts is to subresolve the input particle grain sizes and weight them according to a Dohnanyi distribution; however, this would increase noise properties, and given that our investigation focuses on measuring the noise contribution of exozodi, this is not a viable option. Instead, we opt to subresolve the grain sizes by interpolating over the coarse grain size list, equalizing the weight of the individual grains to maintain the original cross-sections. We subresolve the coarse grain size list into 500 equally spaced grain sizes in log space.  While this is the best option for the present study, which focuses on broadband imaging, it may create a disk color that is redder than that expected from a Dohnanyi size distribution and Mie theory. 

Although using the subresolved and normalized particle size list reduces the Mie theory ringing artifacts, these changes increase the run time of an individual \texttt{dustmap} call by a factor of six, as more Mie theory calculations are required to accommodate the additional grain sizes. 
To reduce runtime while maintaining the reduction of Mie theory ringing artifacts, we limit our subresolving methods to grains $<$ 9.4 $\mu m$ in size. 
The fractional difference between the contrast curves calculated with the partially-subresolved grain size list and the fully-subresolved list is  $< 0.1 \%$ (plotted in the bottom panel of Figure~\ref{fig:particlesize}). 
The final run time for an individual \texttt{dustmap} call using the partially-subresolved grain size list is a factor of 1.7 slower than using the original coarse grain size list and well within tolerance to run on a personal laptop in a few days. 
All \texttt{dustmap} output models are publicly available\footnote{https://asd.gsfc.nasa.gov/Christopher.Stark/catalog.php}.

\subsection{Synthesizing coronagraph observations}

After generating images of our structured exozodiacal disk models, we inject an Earth-like planet and a Sun-like star into the system and convolve the astrophysical scene with simulated spatially dependent point-spread functions (PSFs) for high-contrast coronagraphs.
We simulate PSF subtraction to further suppress the stellar speckles, considering both reference differential imaging and angular differential imaging. 

\subsubsection{Astrophysical scenes}
Our astrophysical scenes are comprised of the exozodiacal images discussed in Section~\ref{sec:disk_sims}, a Sun-like star, and an Earth twin. We manage the star and planet separately from the disk, allowing us to convolve the coronagraph's PSF with the disk while treating the star and planet with individualized on- and off-axis PSF models, and add all sources together later. The systems are placed 10 pc from Earth. We assume the host star has 1 R$_\odot$, 1 L$_\odot$, T$_\mathrm{surface}$ $= 5770$ K, with a magnitude of 4.83 in the V-band and an angular diameter of 0.465 mas at 10 pc. The planetary companion is an Earth-twin located at quadrature (maximum apparent separation). The planet properties include 1 R$_\oplus$, 1 M$_\oplus$, and an Earth-like albedo derived from models of disk-integrated flux presented in \citet{Stark2022}.

\subsubsection{Coronagraph and PSF models}\label{sec:coro}
We simulate realistic observations of a future high contrast imaging space telescope using the high-contrast coronagraph models described in \citet{Kammerer2022-vh}. We investigate two coronagraph designs, each paired with a mirror that has a different circumscribed diameter. 

The first coronagraph--mirror configuration we consider is an 8 m primary mirror with a deformable mirror-assisted vortex charge 6 coronagraph originally designed for LUVOIR-B (VC6, \citet{Mawet2010}). This VC6 coronagraph design achieves a raw contrast of $< 10^{-10}$ beyond $\sim 5$ \lamD  separation. 

We also investigate a larger mirror size of 12 m, for which we adopt the apodized pupil Lyot coronagraph (APLC) designed for LUVOIR-A \citep{Aime2002, Soummer2005, StLaurent2018}. The APLC design assumes an 18\% bandwidth achieving a raw contrast of less than $10^{-10}$ beyond $\sim 6$ \lamD separation for sufficiently small stellar angular diameters ($\le 0.5$ \lamD). See \citet{Kammerer2022-vh} for a full description of each coronagraph design we use in this work.
The instrument throughputs for both coronagraph cases are assumed to be an unrealistic 100\%--- ultimately this assumption does not matter, as we do not attempt to calculate realistic absolute exposure times in this study, and instead compare the planet's measured S/N to the expected S/N.

We convolved each exozodi model with the spatially varying coronagraph PSF using the coronagraph simulation tool developed by \citet{Kammerer2022-vh}. Briefly, this tool loads a pre-generated discrete set of off-axis PSFs, interpolates them to form a 3D datacube of PSFs centered on each pixel, and then performs a fast matrix multiplication to convolve with the astrophysical scene, creating realistic simulated direct imaging observations. 

The planet's PSF is modeled by simply interpolating the discrete set of off-axis PSFs over position on the detector, and evaluating at the radial offset and position angle of the planet.  

For the star, we use a pre-generated discrete set of on-axis PSFs calculated over a range of stellar angular diameters. We interpolate this set of PSFs over stellar diameter, and evaluate at the angular diameter of 0.465 mas, which is the size we assume for the case of a Sun-like star at 10 pc away from Earth. 

\subsubsection{Photon noise}

After convolving the images by the PSF models described in Section~\ref{sec:coro}, we scale the images to a constant detector pixel scale of $0.5$ \lamD for an observing wavelength of $0.5 \mu m$ and add photon noise. In this work, we do not consider nor model detector noise. We adopt an exposure time sufficient to set a planetary S/N of 7 (see Section~\ref{sec:comparison}), and calculate the number of photons collected per pixel on the detector. To add photon noise to each pixel, we draw from a Poisson distribution with a mean corresponding to the number of photons collected in the noiseless pixel.

\subsubsection{PSF subtraction}\label{sec:PSF_sub}
Although the stellar light is suppressed to a raw contrast of $\sim 10^{-10}$, the optics leave behind a field of spatially variable residuals known as speckles. To further suppress these speckles, we assume that two images are observed: a science and a reference image. The science image is always an image of the target system, while the reference image can either be another image of the science target, or an observation of a similar, but isolated, stellar target. Subtracting the science and reference images allows us to suppress the speckle field. 

We consider two possible methods of PSF subtraction: reference differential imaging (RDI) and angular differential imaging (ADI). The RDI technique removes the residual stellar speckle pattern by empirically measuring the speckle field of an isolated, but otherwise similar star to the science target. This image is subtracted from the science image to remove the speckle field. For RDI, we make the ideal assumption that the reference star is identical to the science target. 

The ADI technique removes the residual stellar speckle pattern by using two images of an astrophysical scene separated a roll angle. Because the optics internal to the observatory roll with the telescope, the speckle pattern remains stationary on the detector, while the astrophysical scene is rotated according to the roll angle. By taking two exposures at different roll angles, a target can therefore serve as its own reference star, though this results in some degree of disk self-subtraction, as well as positive and negative copies of planetary companions. For ADI, we assume a $30^\circ$ roll angle, which is approximately the minimum roll angle required to avoid planetary self-subtraction for a system 10 pc away with a planet at orbital radius of 1 AU. 

If our telescope were perfectly stable, the two speckle patterns would subtract to the Poisson noise limit. In reality, time-varying wavefront error (WFE) will result in two slightly different speckle patterns, such that the PSF subtraction is imperfect and we are left with a systematic noise floor. We included this effect by adopting unique WFE time series for the science and reference observations. These WFE time series are propagated through the coronagraph model as optical path difference (OPD) error maps present at the entrance pupil of the coronagraph that vary as a function of time (during 20 seconds, corresponding to 8000 OPD maps). These time series for the 8 m \citep{Potier2022_LUVB} and 12 m \citep{Potier2022_LUVA} designs were generated by Lockheed Martin via an integrated model of the telescope and spacecraft structural dynamics, and include the rigid body motion of the primary mirror segments, the dynamic interaction of flexible structures, and the disturbances from the pointing control system. The two WFE time series produce speckle fields that differ by $<$ 1 \% of the raw contrast.

\subsection{Exozodi mitigation and planet detection}

Regardless of the PSF subtraction technique used, exozodiacal disk structure does not fully self-subtract and significant residual structure is left in the subtracted image (see left panel of Figure~\ref{fig:structure_hipass}). Because this structure is spatially inhomogeneous, it cannot be easily removed with high-order polynomials. We thus convolve the image with a high-pass filter to model and subtract the disk structure from the image. To detect the planetary signal in the disk-subtracted image, we use either aperture photometry or a PSF matching technique. Finally, we measure the noise in a region immediately surrounding the planet, and compute the planetary S/N.

\subsubsection{Exozodi subtraction via high-pass filtering}\label{sec:hipass}
To remove the exozodiacal disk and its structure, we convolve our synthesized observations with a 2D Gaussian high-pass filter. For each scenario, we optimize the FWHM of the filter to remove the residual disk structure while preserving the planet signal by applying filter sizes ranging from 0.5 to 50 \lamD to the image, and choosing the filter size that maximizes the measured planetary S/N.  A filter size identical to the size of the image (in our case, 50 \lamD) effectively does nothing, while a filter size of 0.5 \lamD, identical to the pixel scale, removes all information, including point sources. 

\begin{figure}
    \centering
    \includegraphics{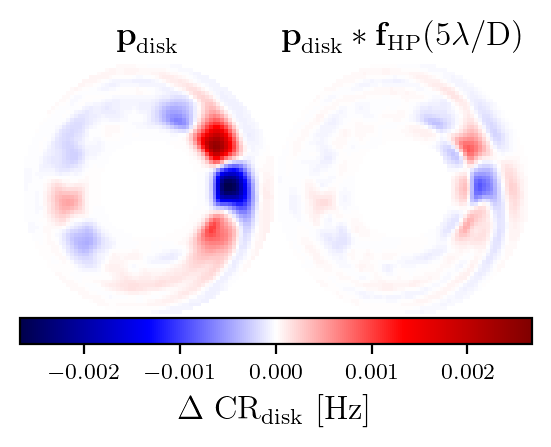}
    \caption{ADI PSF subtraction for an 8 m mirror configuration of a 1 zodi face-on exozodiacal disk, $\mathrm{\textbf{p}}_\mathrm{disk}$, before (left) and after (right) convolution with a 5 \lamD high-pass filter, $\mathrm{\textbf{f}}_\mathrm{HP}(5 \lambda / \mathrm{D})$). }
    \label{fig:structure_hipass}
\end{figure}

Generally, exozodiacal structure is an extended source, and a filter size comparable to the physical size of the exozodiacal structure will ideally remove residual structure in a subtracted image. The minimum filter size we consider is set by the instrumental PSF: a filter size comparable to the size of the PSF will subtract planetary signal. Figure~\ref{fig:structure_hipass} shows a pre- and post-processed image of an ADI subtracted face-on exozodiacal disk structure ($\mathrm{\textbf{p}}_\mathrm{disk}$) convolved with a high pass filter ($\mathrm{\textbf{f}}_\mathrm{HP}$) with a filter size of 5 $\lambda / D$ for an 8 m mirror configuration.  

\begin{figure*}
    \centering
    \includegraphics{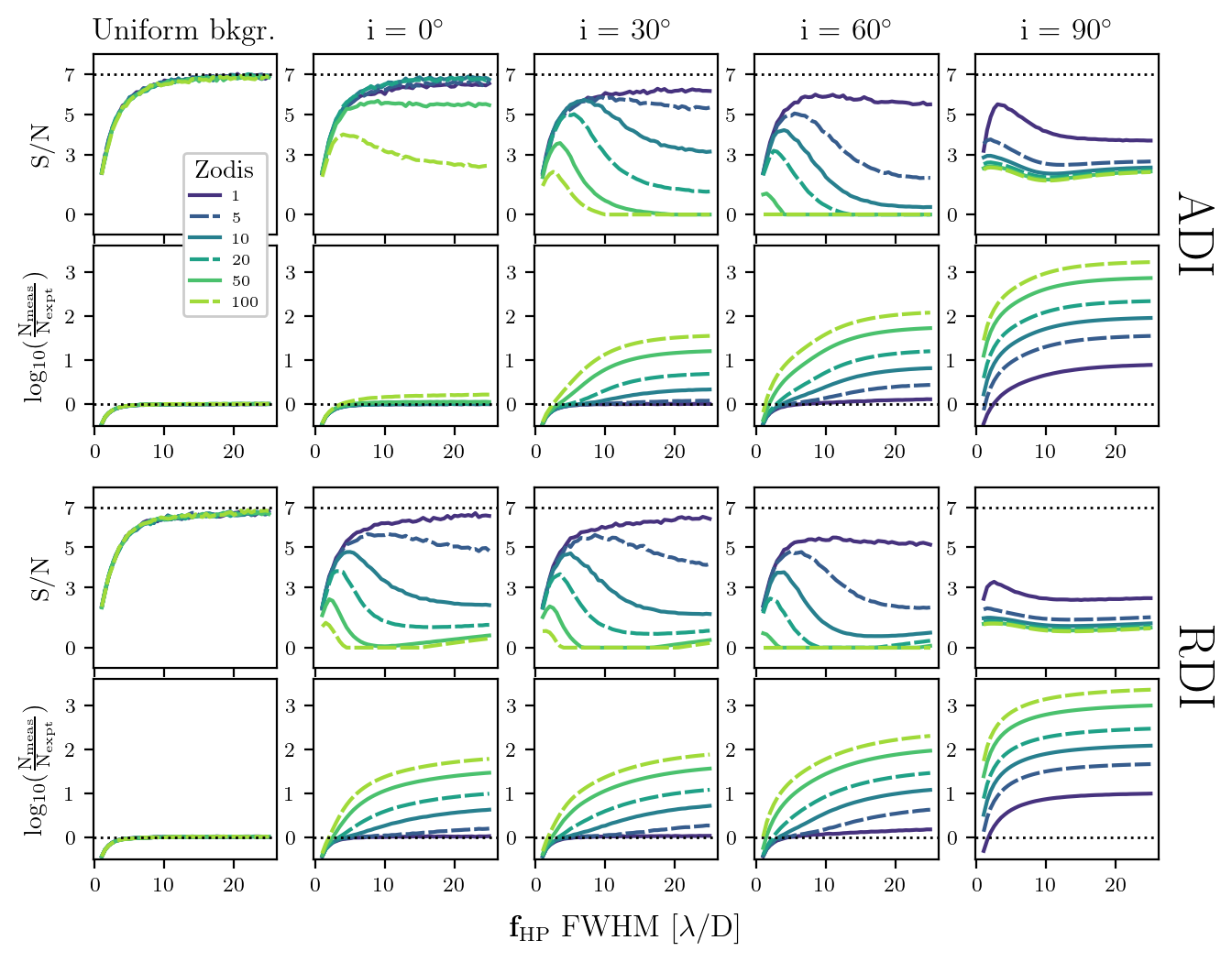}
    \caption{Measured planetary S/N (first and third rows) and measured noise relative to the expected Poisson noise (second and fourth rows) as a function of the applied high-pass filter size for an 8 m mirror configuration. The high-pass filter size is the FWHM of a two-dimensional Gaussian in units of \lamD convolved with ADI (upper two rows) and RDI (lower two rows) PSF subtracted images. Each panel includes all zodi levels considered in this work. The horizontal dotted line represents either the input planetary S/N or the Poisson noise limit. For most cases, an optimal high-pass filter size exists when the measured noise in the filtered image is equal to the expected Poisson noise; the planetary S/N measured in the optimally filtered image is our fiducial S/N measurement.}
    \label{fig:SNR_noise_vs_HPFS}
\end{figure*}

For each combination of inclination and zodi level, we vary the size of the high-pass filter in increments of 0.5 \lamD, and measure the noise after the filter is applied (see Section~\ref{sec:measure_SNR} for a description of the noise measurement technique).  Figure~\ref{fig:SNR_noise_vs_HPFS} shows both the measured S/N of the planet and the ratio of measured to expected noise as a function of the filter size in units of $\lambda / D$ for all zodi levels at all inclinations for both ADI and RDI PSF subtraction methods for an 8 m mirror configuration. We also include a case with a uniform disk background for comparison. 

We then identify the optimal filter size for a given simulation that reduces the measured noise in the image to the expected Poisson noise limit, thereby maximizing the measured planetary S/N. For low zodi, low inclination cases, filter sizes of $\sim 20$ $\lambda / D$ (40 pixels) are sufficient for removing the disk structure. High zodi, high inclination cases require smaller, more aggressive filter sizes which consequently also subtract planetary signal. Furthermore, we are unable to remove the disk structure down to the Poisson noise limit for edge-on cases with $>1$ zodis using any filter size. Applying a small enough filter size to remove disk structure down to the Poisson noise limit coincides with the maximum planetary S/N we are able to measure (see Figure~\ref{fig:SNR_noise_vs_HPFS}), and we choose this optimal filter size for each simulation we consider. We report the filter size used for each scenario in Table~\ref{tab:optimal_fs}.

\begin{deluxetable*}{cc|cccc|cccc}\label{tab:optimal_fs}
\tablecaption{Chosen filter sizes for each system inclination and zodi level optimized for subtracting exozodi to the Poisson noise limit. The upper and lower quadrants of this table correspond to ADI and RDI PSF subtraction, respectively. The left and right quadrants correspond to 8 m and 12 m mirror architectures.  }
\tablewidth{0pt}
\tablehead{ 
\colhead{} & \colhead{} & \multicolumn{4}{c}{\large 8 m mirror} & \multicolumn{4}{c}{\large 12 m mirror}  \\
\colhead{} & \colhead{zodis} & $0^\circ$ incl. & $30^\circ$ & $60^\circ$& $90^\circ$ & $0^\circ$ & $30^\circ$ & $60^\circ$ & $90^\circ$ 
}
\startdata
\multirow{6}{*}{\rotatebox[origin=c]{90}{{\large ADI}}}  & 1 & 30 & 46 & 28 & 8 & 20 & 45 & 13 & 5 \\ 
 & 5 & 50 & 18 & 11 & 4 & 29 & 18 & 10 & 3 \\ 
 & 10 & 34 & 13 & 8 & 3 & 35 & 13 & 7 & 2 \\ 
 & 20 & 29 & 9 & 6 & 2 & 27 & 10 & 6 & 2 \\ 
 & 50 & 17 & 6 & 3 & 2 & 11 & 7 & 4 & 2 \\ 
 & 100 & 12 & 4 & 3 & 2 & 7 & 5 & 3 & 2 \\ 
 \hline
\multirow{6}{*}{\rotatebox[origin=c]{90}{{\large RDI}}}  & 1 & 11 & 12 & 9 & 5 & 19 & 14 & 10 & 4 \\ 
 & 5 & 10 & 11 & 8 & 3 & 11 & 11 & 8 & 2 \\ 
 & 10 & 7 & 7 & 6 & 2 & 9 & 8 & 7 & 2 \\ 
 & 20 & 5 & 5 & 4 & 2 & 7 & 7 & 5 & 2 \\ 
 & 50 & 4 & 3 & 3 & 2 & 5 & 5 & 3 & 2 \\ 
 & 100 & 3 & 3 & 2 & 2 & 4 & 4 & 2 & 2 \\ 
\enddata
\end{deluxetable*}

\subsubsection{Measuring SNR via aperture photometry}\label{sec:ap_phot}

We measure the signal of the planet and the noise in the region surrounding the planet by placing apertures of radius 0.7 \lamD, which is approximately the size of the planetary PSF, and summing the signal within each aperture. Because the background of the image, including exozodi and PSF subtraction residuals, is spatially inhomogenous (especially for inclined systems), noise estimation is sensitive to both the location and size of the region used to sample the noise. Thus, we are often limited in the number of resolution elements we can use to estimate the noise, and we adopt the small sample statistics formalism recommended by \citet{Mawet2014}. We calculate S/N by employing the two-sample t-test to determine the significance of one resolution element (i.e. the signal) compared to the resolution elements in a given region of the image (i.e. the area used to estimate noise). We use the following equation to calculate planetary S/N:

\begin{equation}\label{eq:SNR}
    \mathrm{S/N} = \frac{x_{00}}{\sigma_n \sqrt{1 + \frac{1}{N_n}}} 
\end{equation}

In Equation~\ref{eq:SNR}, $x_{00}$ is the intensity of the planet signal, and $\sigma_n$ is the standard deviation of the resolution element intensities used for noise estimation with $N_n - 1$ degrees of freedom, where $N_n$ is the number of resolution elements used for noise estimation. The term $\sqrt{1 + \frac{1}{N_n}}$ is a correction factor for small number statistics as derived in the two sample t-test formalism of \citet{Mawet2014}. The noise term $\sigma_n$ is defined as
\begin{equation}\label{eq:noise}
    \sigma_n = \sqrt{\frac{\sum (x_{ij} - \bar{\bm x})^2}{N_n - 1}},
\end{equation}
where $x_{ij}$ is a resolution element intensity centered on pixel (i,j) calculated within a defined region suitable for estimating the noise at the planet location, and $\bar{\bm x}$ is the mean of an array of resolution element intensities $x_n$.

Assuming the location of the planet is known, we measure the planet signal, $x_{00}$, by placing an aperture of radius 0.7 \lamD centered on the planet location and summing the pixel values within the aperture. In the case of ADI PSF subtraction, which contains a ``positive'' and a ``negative'' copy of the planet separated by the roll angle, we place an additional aperture on the ``negative'' copy of the planet in the PSF subtracted image. The sum of the absolute values of the ``negative'' planet signal and the ``positive'' planet signal is the total planetary signal in the image.

\begin{figure*}
    \centering
    \includegraphics{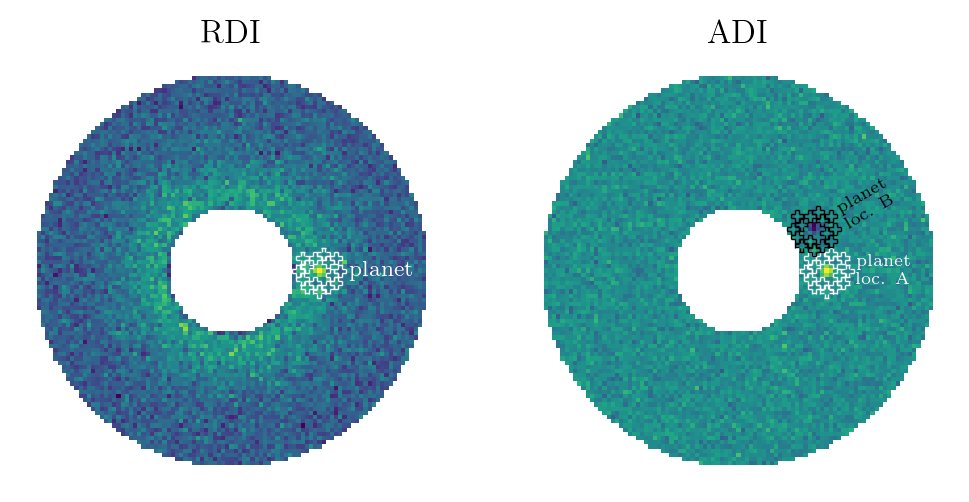}
    \caption{Schematic showing the regions used to measure noise for RDI (left) and ADI (right) PSF subtracted images for an 8 m mirror configuration.}
    \label{fig:noise_measurement}
\end{figure*}

Because an inclined exozodiacal disk can exhibit significant forward scattering and is not azimuthally symmetric, we measure noise in a local region immediately surrounding the planet. We define a small annulus centered on the planet location with an inner radius of 1 \lamD and an outer radius of 3 \lamD, and place apertures within this region. Figure~\ref{fig:noise_measurement} shows a schematic for the noise regions used in RDI and ADI PSF subtraction images for an 8 m mirror configuration. The size of the annulus region limits the number of apertures we can place; however, using the ADI PSF subtraction technique allows us to double the number of apertures we place because two copies of the planet exist. We therefore place $\sim 15$ and $\sim 30$ apertures in RDI and ADI subtracted images, respectively. We sum the intensities within each aperture and create an array of noise measurements, $\bm x$. The standard deviation of the array of noise measurements is calculated using Equation~\ref{eq:noise}, and this value multiplied by the correction factor in the denominator of Equation~\ref{eq:SNR} is the measured noise. 

\subsubsection{Measuring SNR via matched filtering}\label{sec:measure_SNR}
In addition to measuring S/N using aperture photometry, we also consider a more advanced PSF matching technique for measuring planetary S/N. The PSF matching technique leverages the known shape of an off-axis PSF via matched filtering to detect possible point source companions more robustly than aperture photometry \citep[e.g.][]{Kasdin2006}. In this approach, we use our library of off-axis coronagraphic PSF models described in Section~\ref{sec:coro} to interpolate an offset PSF model centered on each pixel in the image. For each pixel, the PSF model is then truncated (i.e. everything outside of some radius is set to zero), and the truncated model is normalized. We explored a range of truncation radii, and found a radius of 0.7 \lamD was sufficient for our purposes. In this matched filtering formalism, the intensity of a given resolution element centered on location (i, j) is given by 
\begin{equation}\label{eq:signal}
    x_{ij} = \frac{(\bm{p} \ast \bm f_{\mathrm{HP}}) \cdot \bm m_{ij}}{ \bm m_{ij} \cdot \bm m_{ij} }, 
\end{equation}
where $\bm{p}$ is the vectorized PSF-subtracted image, $f_{\mathrm{HP}}$ is the Gaussian high pass filter, $m_{ij}$ is the vectorized matched filter PSF model for pixel (i, j), and $\ast$ and $\cdot$ are the convolution and dot product operators, respectively. The intensity of the planetary signal, $x_{00}$, is thus Equation~\ref{eq:signal} applied at the planet location. We estimate the noise using the technique described for aperture photometry; however, instead of placing apertures within the defined noise region, we apply Equation~\ref{eq:signal} to the same locations as in aperture photometry.

The above describes a generalized matched filtering formalism which can be directly applied when using the RDI PSF-subtraction technique. However, for the ADI technique, the planet signal appears as two components in the subtracted image: a ``positive'' and a ``negative'' signal separated by the roll angle. In this case, we modify our matched filter PSF model slightly by appropriately including a negative PSF companion in the $\bm m_{ij}$ term of Equation~\ref{eq:signal}, separated by the roll angle. We multiply the companion PSF by $-1$, and in a similar procedure as the generalized formalism, we truncate this companion PSF to a radius of 0.7 \lamD and normalize the entire PSF model such that its absolute sum is unity. The ``negative'' PSF copy is offset from the ``positive'' one by accounting for the roll angle of the telescope, and this offset changes with angular separation from the star.  This results in a unique matched filter PSF model for each pixel (i,j). The PSF pair is fixed by the roll angle, and therefore change in separation with circumstellar distance. This may help to reduce the impact of stellar speckles, though we do not investigate this effect here.   When this matched filter is convolved with an image, the contributions from both the ``positive'' and ``negative'' images are included in the calculation for the intensity of a given resolution element (Equation~\ref{eq:signal}).  

\subsection{Comparing to expected values}\label{sec:comparison}
After post-processing the synthesized images, we compare our measurements for planetary S/N and photon noise to their expected values. We adopted an exposure time sufficient to set a planetary S/N of 7, given by
\begin{equation}\label{eq:tint}
    T_{\mathrm{int}} = (\mathrm{S/N})^2 \frac{\mathrm{CR}_\mathrm{plan} + \mathrm{CR}_\mathrm{back}}{\mathrm{CR}_\mathrm{plan}^2}, 
\end{equation}
where $T_{\mathrm{int}}$ is the total integration time of the observation, $\mathrm{SNR}$ is the signal-to-noise ratio of a faint planet companion, $\mathrm{CR}_\mathrm{plan}$ is the photon count rate of the planet, and $\mathrm{CR}_\mathrm{back}$ is the photon count rate of the background count rate consisting of stellar speckle and exozodiacal disk contributions. The photon count rates are calculated by integrating over an aperture of radius 0.7 \lamD in the noiseless images. The stellar and disk contributions of $\mathrm{CR}_\mathrm{back}$ vary according to the PSF subtraction method (see Section~\ref{sec:PSF_sub}), and are given by 
\begin{equation}\label{eq:background}
    \mathrm{CR}_\mathrm{back} = 
    \begin{cases}
        2 \mathrm{CR}_\mathrm{star} + \mathrm{CR}_\mathrm{disk} & \text{for RDI} \\
        2 \mathrm{CR}_\mathrm{star} + 2 \mathrm{CR}_\mathrm{disk} & \text{for ADI}
    \end{cases},
\end{equation}
where $\mathrm{CR}_\mathrm{star}$ and $\mathrm{CR}_\mathrm{disk}$ are the average count rates of the stellar speckles and exozodiacal dust in the region used to ultimately measure noise in the post-processed image (see Section~\ref{sec:measure_SNR}). In Equation~\ref{eq:background}, $\mathrm{CR}_\mathrm{star}$ is doubled for both RDI and ADI because we assume that we spend the same amount of time integrating on the reference image as we do on the science image. Only one $\mathrm{CR}_\mathrm{disk}$ term is counted for RDI because we assume that the reference star is a perfect copy of the star in the science image and does not include a debris disk or planets. 

In the following section, we report planetary S/N and photon noise measurements using both aperture photometry and PSF matching, and compare these values to the input S/N and the expected background noise given by Equation~\ref{eq:background}. We report these relative comparisons for all zodi, disk inclination, primary mirror sizes, and PSF subtraction techniques considered in this work.

\section{Results}\label{sec:results}
Using the tools and models described in the previous section, we synthesize realistic high-contrast observations of Earth-like exoplanets in systems with significant exozodiacal disk structure. We remove exozodiacal disk structure from the observations by applying an optimized high-pass filter, and compare the measured residual noise in the post-processed image to the expected Poisson noise. The process of simulating observations, removing exozodi, and measuring noise and S/N is repeated 1000 times for each zodi level/inclination configuration to average over photon noise, and we report our measurements as the median value of the iterations for each configuration. Unless otherwise specified, results are reported primarily for the 8 m mirror configuration to provide estimates for the nominal $\sim 6$ m inscribed mirror recommendation of the \citet{Astro2020}.

\subsection{Disk Calibration}
The input disk models have inherent noise associated with the limited resolution of N-body simulations (see Section~\ref{sec:dithering}), and we test how this limitation affects our simulations. We attempted to smooth over particle noise by dithering the disk models in both the radial and longitudinal directions, and although our dithering process significantly reduced the particle noise, it did not eliminate this noise source entirely (see Figure~\ref{fig:dither}). We therefore consider how particle noise contributes to the overall noise.  

\begin{figure*}
    \centering
    \includegraphics{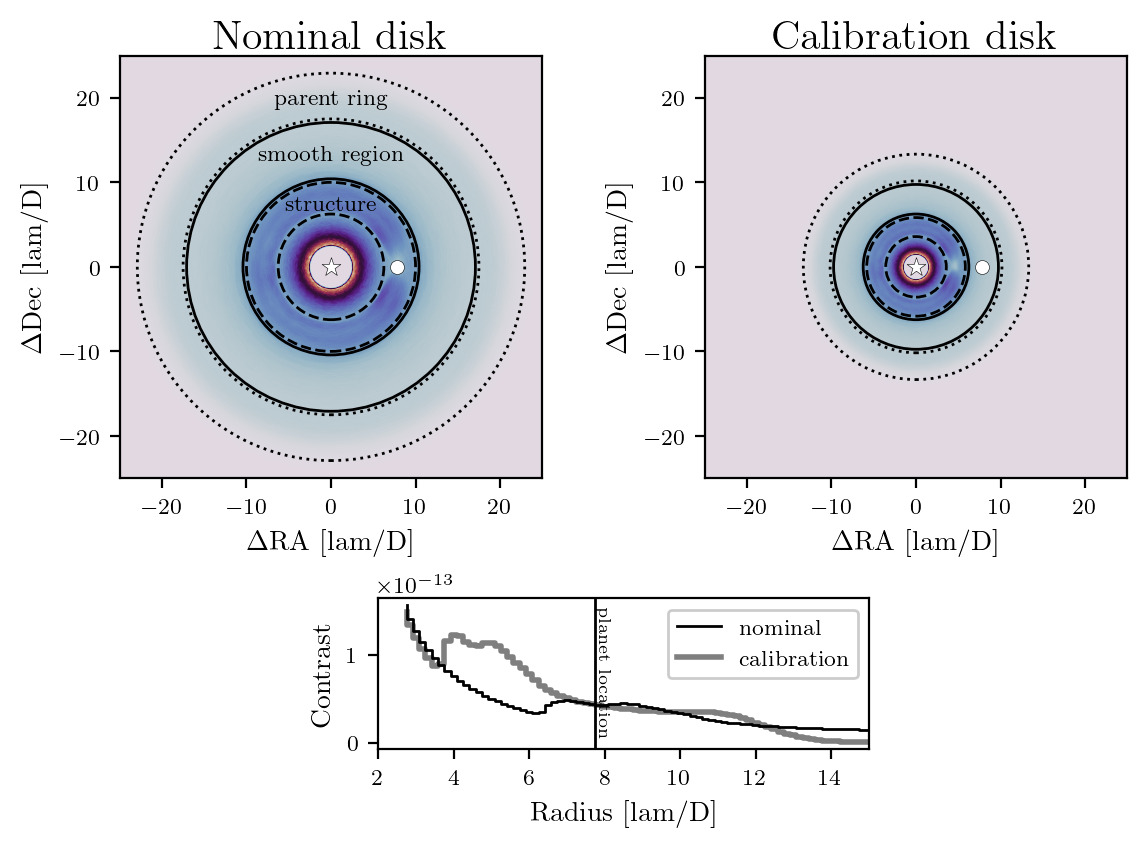}
    \caption{Comparison of the nominal disk model and the calibration disk model assuming observations with an 8 m mirror architecture. The planet in the nominal system is located in a region of disk structure, while the planet in the calibration disk model has a smoothly varying background. The lower panel shows azimuthally averaged disk-to-star contrast as a function of radius for the nominal and calibration disks.}
\label{fig:struct_vs_calibrsmooth}
\end{figure*}

To test whether particle noise in the disk model significantly contributes to the noise budget, we re-scale the nominal disk models to construct calibration disk models that do not exhibit mean-motion resonant structure at the orbital radius of the planet.  The exozodiacal debris disks in this work are generally composed of the three distinct regions shown in Figure~\ref{fig:struct_vs_calibrsmooth}: a parent ring where the dust particles originate, mean motion resonant structures at the orbital radius of the planet, and smoothly varying dust between the parent ring and structure. The disk models in Figure~\ref{fig:struct_vs_calibrsmooth} include the inverse square illumination factor, but the true calibration models do not include scattered stellar light to help isolate the source of the particle noise. We construct the calibration disk models by re-scaling the size of the disk model such that the planet lies in the center of the structure-less ``smooth region''. Subsequently, we re-scale the contrast of the new calibration disk to match the contrast of the nominal disk at the planet location. Figure~\ref{fig:struct_vs_calibrsmooth} illustrates this process for an 8 m mirror configuration.

\begin{figure}
    \centering
    \includegraphics[]{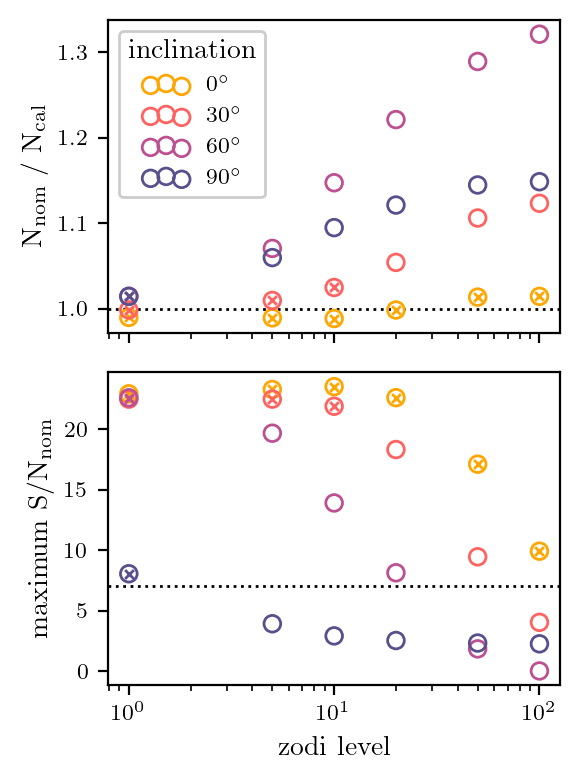}
    \caption{Upper panel: Ratio of measured noise in the nominal disk model to calibration disk model as a function of zodi level. Markers within $\pm 5\%$ of the horizontal dotted line (with an ``x'' in the center) indicate cases where the total measured noise in the nominal disk model is dominated by particle noise inherent to the N-body models. Lower panel: Maximum measurable S/N for cases that include the nominal disk model. For most edge-on or high-zodi cases, the input S/N of 7 (dotted horizontal line) is not recoverable due to systematic noise associated with the physical disk properties. The results in this figure assume an 8 m primary mirror and ADI PSF subtraction. }
    \label{fig:calib_nom_noise}
\end{figure}

We insert the calibration disk models into astrophysical scenes in lieu of the nominal disk models, and apply the same treatment described in Section~\ref{sec:methods}, assuming an exposure time given by Equation~\ref{eq:tint}. We plot the ratio of measured noise in the nominal disk model to that of the calibration disk model for each inclination as a function of zodi level in the upper panel of Figure~\ref{fig:calib_nom_noise}. We consider noise ratios within $\pm5\%$ of unity to be limited by the particle noise, and indicate these cases with a small ``x'' in the center of their markers.  Cases limited by particle noise include $0^\circ$ inclination for all zodi levels, $30^\circ$ inclination with 1, 5, and 10 zodis, $60^\circ$ inclination with 1 zodis, and $90^\circ$ inclination with 1 zodis. All other cases are limited by the physical properties of the disk, including structure and spatially-varying brightness. However, we do not expect the cases limited by particle noise to impact the validity of our analysis, as explained below. 

Systematic noise from either N-body particles or disk structure can limit the maximum recoverable planetary S/N, but all cases limited by particle noise have maximum S/N values greater than our target S/N. To calculate this maximum S/N limitation for each of our scenarios, we run our analysis pipeline adopting an exposure time of $10^{10}$ s, which approximates an observation with infinite exposure time, allowing the measured planetary S/N to saturate at the maximum possible value. We assume an 8 m mirror and ADI PSF subtraction. We report these values in the lower panel of Figure~\ref{fig:calib_nom_noise}, and flag the scenarios dominated by particle noise. All cases limited by particle noise have maximum S/N measurements greater than our input planetary S/N of 7. We therefore conclude that the particle noise inherent to the N-body simulations will not affect the validity of our results. 

Some maximum S/N measurements for high-inclination, high-zodi cases are less than our input planetary S/N of 7; these cases include $30^\circ$ inclination with 100 zodis, $60^\circ$ inclination with $\geq 50$ zodis, and $90^\circ$ inclination with $\geq 5$ zodis (see Figure~\ref{fig:calib_nom_noise}). However, these cases are not dominated by particle noise, and instead are limited by the disk structure.  For high-inclination, high-zodi cases, it is impossible to recover the input S/N due to systematic noise associated with the disk structure, even with unlimited exposure time. However, this is a limitation of our methodology, and not an absolute limitation. Other techniques may be required to mitigate the disk in these examples (see Section~\ref{sec:systematics} for further discussion).

\subsection{Planet detection and S/N measurements}\label{sec:results_SNR_meas}
\begin{figure*}
    \centering
    \includegraphics{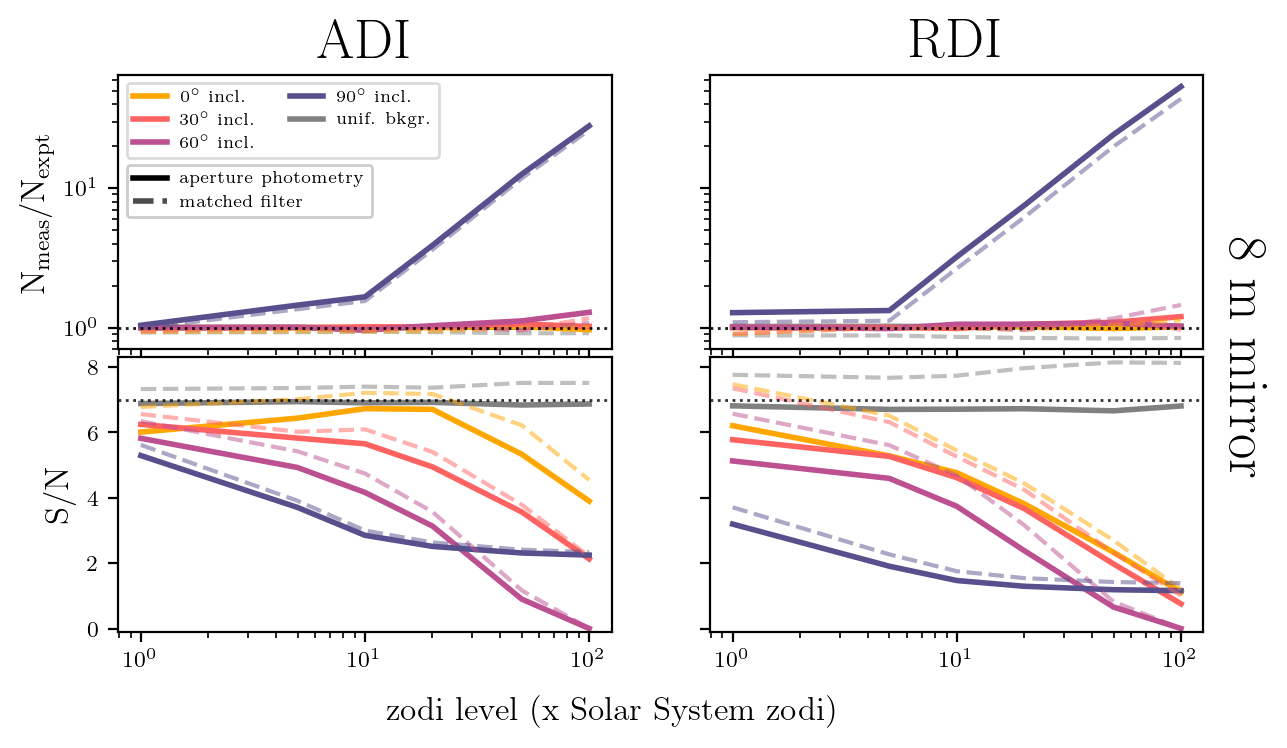}
    \caption{Measured noise relative to the expected Poisson noise (top row) and measured planetary S/N (bottom row) as a function of the zodi level of a system after optimizing the high-pass filter size to maximize S/N for an 8 m mirror architecture. The columns are results for ADI and RDI PSF subtraction techniques. Each panel includes results using aperture photometry and PSF matching, designated by solid and dashed lines, respectively. Each panel shows results for all system inclinations considered in this work, as well as systems with uniform backgrounds for comparison. We find that applying a high-pass filter can subtract exozodiacal disk structure to the Poisson noise limit at the expense of signal for nearly all systems $<90^\circ$ inclined; however edge-on systems remain challenging.}
    \label{fig:SNR_vs_zodis}
\end{figure*}

\begin{figure*}
    \centering
    \includegraphics{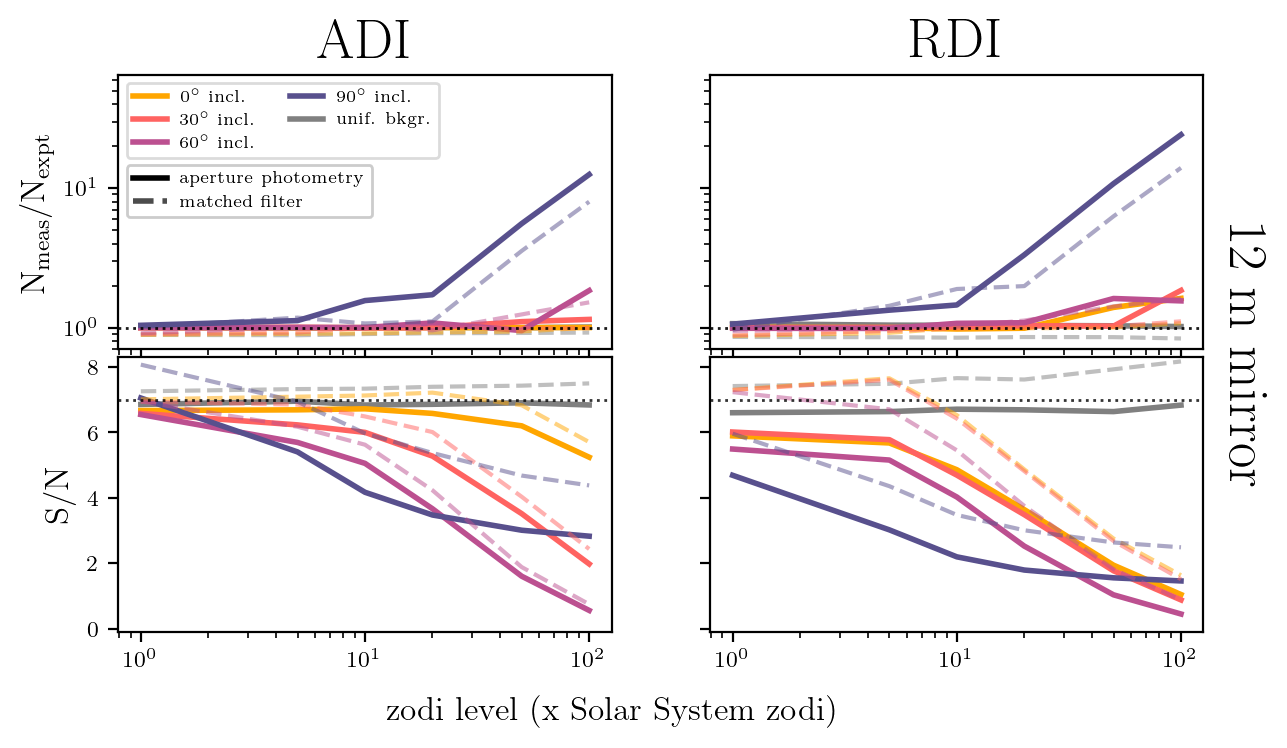}
    \caption{Same as Figure~\ref{fig:SNR_vs_zodis}, but for a 12 m mirror architecture. The smaller PSF of the larger mirror size helps to ``resolve out'' the exozodi, and thus the measured planetary S/N is improved in nearly all scenarios.}
    \label{fig:SNR_vs_zodis_LUVA}
\end{figure*}

After applying our high-pass filtering routine to subtract residual disk structure from PSF-subtracted images, we measure both the signal of the planet and an estimate for the noise in the post-processed image using the formulae and processes described in Sections~\ref{sec:ap_phot} and~\ref{sec:measure_SNR}. Here we report both the recovered S/N of the exoplanet, and the estimated noise at the planet location as it compares to the expected Poisson noise ($\mathrm{CR}_\mathrm{back}$) given by Equation~\ref{eq:background}. The planet was injected in the image at a S/N of 7, and the integration time for each case was calculated using Equation~\ref{eq:tint}. Figures~\ref{fig:SNR_vs_zodis} and~\ref{fig:SNR_vs_zodis_LUVA} show recovered planetary S/N measurements in the ADI and RDI PSF-subtracted post-processed image as well as the ratio of measured (N$_\mathrm{meas}$) to expected (N$_\mathrm{expt}$) noise in the image for a space-based direct imaging telescope with 8 m and 12 m architectures, respectively, for all cases. In these plots, N$_\mathrm{expt}$ is the expected background noise at the location of the planet, and is calculated by multiplying the count rate of the background noise given by Equation~\ref{eq:background} by the exposure time we adopt (Equation~\ref{eq:tint}). We also include results for a system with a uniform, completely smooth exozodi background for comparison.  Figures~\ref{fig:SNR_vs_zodis} and~\ref{fig:SNR_vs_zodis_LUVA} also present a comparison of planet detection methods, with aperture photometry (Section~\ref{sec:ap_phot}) and PSF matching (Section~\ref{sec:measure_SNR}) plotted as solid and dashed lines, respectively. 

The noise vs. zodi panels of Figures~\ref{fig:SNR_vs_zodis} and ~\ref{fig:SNR_vs_zodis_LUVA} suggest that it is possible to choose a high-pass filter size that subtracts exozodiacal dust structure down to the Poisson noise limit in nearly all cases, except for edge-on systems with $> 1$ zodi, for both ADI and RDI PSF subtraction routines. However, the S/N vs. zodi panels show that in some cases we are unable to recover the input planetary S/N in the post-processed image with the integration time specified by Equation~\ref{eq:tint}. For high-zodi, high-inclination systems, the maximum measurable S/N due to systematic noise (see Figure~\ref{fig:calib_nom_noise}) is lower than the input S/N. 

We note a peak at $\sim 10$ zodis in the S/N curve for the face-on ($0^\circ$ inclined) ADI case of Figure~\ref{fig:SNR_vs_zodis}. As described in Section~\ref{sec:hipass}, filter size selection can impact the planetary signal, and we optimize the balance between exozodi subtraction and planetary signal preservation by choosing the high pass filter size that maximizes the planetary S/N (see Table~\ref{tab:optimal_fs}). In the low-zodi regime of the face-on ADI S/N curves, the optimal filter size is larger than the scale of the mean motion resonance structure because this maximizes the measured S/N. However, in these cases the structure is not fully mitigated and the corresponding noise term remains, slightly reducing the measured S/N. This effect is less pronounced at a density of $\sim 10$ zodis because the structure accounts for a smaller percentage of the overall surface brightness in cases with increased disk density due to enhanced collisional destruction of grains in denser disks, resulting in an overall smoother disk profile amenable to efficient removal via our high-pass filter technique. 

The choice of PSF subtraction technique also affects our S/N measurements. In Figures~\ref{fig:SNR_vs_zodis} and~\ref{fig:SNR_vs_zodis_LUVA}, results for ADI and RDI PSF-subtracted images are presented in the left and right columns, respectively. In both cases, the input planetary S/N is able to be recovered with a uniform disk background; however, introducing disk structure into a system results in clear differences between the measured S/N values in the ADI and RDI cases. For face-on cases, planetary S/N does not degrade until 20 and 5 zodis are reached for the ADI and RDI PSF subtraction techniques, respectively. For inclinations $>30^\circ$, PSF subtraction using either the ADI or RDI technique produces similar trends, although S/N measurements for ADI are up to $\sim 30 \%$ larger than for RDI. We were unable to recover the expected S/N of 7 in these high-inclination cases, even for a cases with a density of 1 zodi. We also compare the aperture photometry and PSF matching techniques for planet detection in Figures~\ref{fig:SNR_vs_zodis} and~\ref{fig:SNR_vs_zodis_LUVA}, and find that using the PSF matching technique yields up to $\sim 10\%$ and $\sim 25 \%$ higher S/N measurements for the 8 m and 12 m cases, respectively. This improvement in S/N translates to $\sim 20\%$ and $\sim 50 \%$ reductions in the required exposure times to achieve the aperture photometry S/N in most cases. 

Figure~\ref{fig:calib_nom_noise} suggests that the presence of systematic noise associated with disk structure may limit the maximum measurable S/N; however, if this maximum measurable S/N is larger than the desired S/N, it may be possible to integrate for longer on the target to achieve the desired S/N. Accounting for all systematic noise terms, we calculate the integration time necessary to achieve a detection significance of 7 for each simulation we consider, and present the results as ratios with the theoretical integration time in Table~\ref{tab:true_tint}, including 8 m and 12 m primary mirror architectures. 

\begin{deluxetable*}{cc|cc|cc|cc|cc}\label{tab:true_tint}
\tablecaption{Comparison of relative integration times for exozodi with structure (this work, C23) and smooth exozodi \citep[K22,][]{Kammerer2022-vh}. Each cell represents the ratio of realistic integration time needed to account for systematics and the theoretical exposure time given by Equation~\ref{eq:tint}. We assume a target S/N of 7 and ADI PSF subtraction. }
\tablewidth{0pt}
\tablehead{ 
\colhead{} & \colhead{Zodis}  & \multicolumn{2}{c}{Face-on} & \multicolumn{2}{c}{30$^\circ$ incl.} & \multicolumn{2}{c}{60$^\circ$ incl.} & \multicolumn{2}{c}{Edge-on} \\
\colhead{} & \colhead{} & C23 & K22 & C23 & K22 & C23 & K22 & C23 & K22 
}
\startdata
\multirow{6}{*}{\rotatebox[origin=c]{90}{{\large 8 m mirror}}} & 1  & 1.2 & 1.0 & 1.3 & 1.0 & 1.7 &  1.0 & 10 &  \nodata \\
 & 5  & 1.2 & 1.0 & 1.5 & 1.0 & 2.6 &  1.0 & \nodata &  \nodata \\
 & 10  & 1.1 & 1.0 & 1.8 & 1.0 & 4.9 &  1.0 & \nodata &  \nodata \\
 & 20 & 1.2 & 1.0 & 2.5 & 1.0 & 76 &  1.2 & \nodata &  \nodata \\
 & 50  & 2.2 & 1.0 & 17 & 1.0 & \nodata &  1.6 & \nodata &  \nodata \\
 & 100  & 10 & 1.0 & \nodata & 1.0 & \nodata &  2.1 & \nodata &  \nodata \\
 \hline
 \multirow{6}{*}{\rotatebox[origin=c]{90}{{\large 12 m mirror}}} & 1  & 1.2 & 1.0 & 1.2 & 1.0 & 1.2 &  1.0 & 1.0 &  \nodata \\
 & 5  & 1.1 & 1.0 & 1.3 & 1.0 & 1.7 &  1.0 & \nodata &  \nodata \\
 & 10  & 1.1 & 1.0 & 1.4 & 1.0 & 2.6 &  1.0 & \nodata &  \nodata \\
 & 20 & 1.1 & 1.0 & 2.3 & 1.0 & 36 &  1.0 & \nodata &  \nodata \\
 & 50  & 1.3 & 1.0 & \nodata & 1.0 & \nodata &  1.2 & \nodata &  \nodata \\
 & 100  & 2.3 & 1.0 & \nodata & 1.0 & \nodata &  1.6 & \nodata &  \nodata \\
\enddata
\end{deluxetable*}

We also see improvements in measured S/N by increasing the diameter of the telescope's primary mirror. Table~\ref{tab:true_tint} provides a direct comparison for the integration time required to detect an Earth-like exoplanet for each simulation. As mentioned in Section~\ref{sec:coro}, these two mirror architectures differ in both size and coronagraph design. The larger mirror has a smaller PSF, effectively allowing it to ``resolve out'' the extended exozodiacal disk and structure and requiring less exposure time to achieve our desired S/N  (see Table~\ref{tab:true_tint}).

\section{Discussion}\label{sec:discussion}

\subsection{Systematic Noise}\label{sec:systematics}

In this work, we have identified two systematic noise terms that contribute to the overall noise budget of the system, impeding planetary S/N measurements. The first is the particle noise inherent to the input N-body simulations. Although we attempt to smooth over this noise in Section~\ref{sec:dithering}, it nevertheless sets a noise floor (see Figure~\ref{fig:calib_nom_noise}).This noise term does not impact the validity of our results, and will not be relevant for real observations. Despite this systematic noise, we are able to recover the input planetary S/N for all affected cases, except for systems with $> 50 $ zodis. Thus, all S/N measurements $< 50$ zodis are largely unaffected by particle noise, and instead may be limited by systematic noise associated with the disk.  

The second systematic noise term we identify is the noise due to exozodiacal disk structure at the location of the planet. Similarly, this term sets a noise floor that limits the measurable planetary S/N. Some cases may have maximum S/N limits below the desired detection significance, as in the case of high-zodi, high-inclination simulations (see Figure~\ref{fig:calib_nom_noise}). In these cases, it is impossible to achieve S/N beyond these limits because we are unable to both subtract the exozodiacal disk structure down to the Poisson noise limit and preserve the planet signal with our analysis pipeline. In particular, the high-pass filter leverages the fact that disk structure is typically more extended and larger in scale than a planetary point source. In high-inclination scenarios as the disk inclination approaches $90^\circ$, the disk itself becomes brighter due to its forward scattering properties, and its spatial scale is reduced to a sharp, knife-edge feature about the same scale as a planetary PSF (see Figure~\ref{fig:dustmap_ims}). In this scenario, the high-pass filter must be applied with an aggressively small FWHM to fit and remove the disk, and consequently the high-pass filter also removes significant planetary signal in the process. We conclude that these high-inclination systems will likely require an alternate technique that removes the disk, but preserves the planetary signal as much as possible. One option may be to fit the edge-on disk shape with a Gaussian, Lorentzian, or other parametric function centered on the disk. Additionally, it may be possible to leverage wavelength-dependent flux of the disk to better remove it from the system, however we leave these options to future work. We note that \citet{Kammerer2022-vh} had similar difficulties removing the disk contribution for edge-on systems with smooth disks using a high-order polynomial--- the high-inclination disk subtraction problem remains an open issue.

\subsection{Mirror size comparison}
In this work, we test 8 m and 12 m mirror configurations, and present the resulting noise and planetary S/N measurements in Figures~\ref{fig:SNR_vs_zodis} and~\ref{fig:SNR_vs_zodis_LUVA}, respectively. The recovered planetary S/N is generally improved by increasing the primary mirror size from 8 m to 12 m. The pixel size of the detector scales with the inverse of the diameter of the mirror, thus the photons of an extended source are spread over more pixels in the 12 m configuration, resulting in the exozodiacal disk being ``resolved out'' with increased mirror size, and improvements in the measured planetary S/N.

The most extreme improvement in recovered planetary S/N by increasing the mirror size is seen in the low-zodi high-inclination cases. In the 1 zodi, edge-on case, the recovered planetary S/N increases by $\sim 25 \%$ when increasing the mirror diameter from 8 m to 12 m. This is due to the smaller PSF of the larger mirror ``resolving out'' the sharp features of the edge-on disk.

\subsection{Relative Integration Time}\label{sec:int_time}

We compare the relative integration time needed for a S/N $=7$ planetary detection for 8 m and 12 m configurations using the ADI PSF subtraction technique in Table~\ref{tab:true_tint}.  Equation~\ref{eq:tint} gives the total theoretical integration time assumed for our simulations, assuming only photon noise and 100\% instrument throughput, and does not take detector noise into account. Thus, any exposure times derived from this work should not be interpreted as absolute. Without the presence of systematics, we would expect the measured and calibration S/N curves to follow the theoretical S/N function given by solving Equation~\ref{eq:tint} for S/N. However, the presence of systematic noise forces the measured S/N curves to deviate from the theoretical S/N curve, eventually plateauing at a S/N maximum where more integration time does not improve the measurement. For most cases with inclinations $< 90^\circ$, this maximum S/N lies above the S/N $=7$ line, implying that more integration time is required to achieve the desired S/N value. High-zodi cases typically require the largest increase in integration time to achieve the desired S/N, by an order of magnitude or more. Therefore, knowledge of a system's zodi level would be necessary to accurately estimate the integration time in a real-world scenario.

\subsection{Comparison to Smooth Disks}
For systems with smooth exozodiacal disks, \citet{Kammerer2022-vh} found that it may be possible to subtract the disk down to the Poisson noise limit using high-order polynomials and no prior information of the system for all cases up to $\sim 10$ and $\sim 50$ zodi for 8 m and 12 m mirror architectures, respectively. For the present study, we find that it may be possible to use high-pass filtering technique to subtract exozodiacal disks with mean motion resonance structure down to the photon noise limit for all cases up to 100 zodi and $< 90^\circ$ inclination. However, as noted in Section~\ref{sec:results_SNR_meas}, achieving this level of disk subtraction for high-zodi, high-inclination cases requires an aggressive high-pass filter that consequently removes planetary signal, and additional exposure time may be necessary to achieve the desired planetary S/N. 

In Table~\ref{tab:true_tint}, we compare relative exposure times calculated in this work to values derived from the results of \citet{Kammerer2022-vh}. Assuming the planetary signal is perfectly preserved, \citet{Kammerer2022-vh} predict that the theoretical exposure time given by Equation~\ref{eq:tint} will be sufficient to detect planets in systems that are $0^\circ$ and $30^\circ$ inclined with up to 100 zodis, and $60^\circ$ inclined with up to 10 zodis for an observatory with an 8 m aperture. Beyond 10 zodis, up to double the theoretical exposure time may be required to detect planets in a $60^\circ$ inclined system. \citet{Kammerer2022-vh} does not present results for edge-on disks. However, systems with exozodiacal disk structure may require more exposure time, with a few times the theoretical value for most cases, and up to an order of magnitude more in the most extreme examples. Therefore, the presence of exozodiacal disk structure will likely impact exposure time for real observations. 

In summary, it may be feasible to subtract exozodiacal dust for low-zodi, moderately inclined systems whether the disk exhibits completely smooth dust or worst-case-scenario mean motion resonance structure, representing two extremes in disk morphology possibilities.

\subsection{Nearby Systems}
This technique shows promise for effectively removing exozodiacal disk structure for the median observed zodi level of nearby systems. The HOSTS survey reported a best-fit median habitable zone zodi level of 3 zodis with a 95\% upper limit of 27 zodis \citet{Ertel2020-kt}. If these nearby systems included an Earth-like planet in their habitable zones as well as exozodiacal structure, we may be able to detect the planetary companion for all inclinations $\leq 60^\circ$ using an 8 m telescope.   Additionally, it may be feasible to subtract exozodiacal structure down to the Poisson noise limit even at the 95\% upper limit zodi level from these systems for inclinations $< 60^\circ$, however more integration time may be required. Although the orientation of the disks in the \citet{Ertel2020-kt} sample were usually unknown, the median inclination of stellar systems with respect to Earth is statistically $60^\circ$ if all systems are randomly oriented. Therefore, we may be able to contend with exozodiacal disks with significant mean motion resonance structure in over half of all nearby targets, so long as disk densities generally follow the distribution found in \citet{Ertel2020-kt}. To access the other half of possible targets, future studies should focus on exozodi mitigation at higher inclinations.

\section{Conclusions}\label{sec:conclusion}

We simulated high-contrast images of Earth-like exoplanets in astrophysical scenes that include significant exozodiacal disk structure, and quantified our ability to subtract mean motion resonant structure to detect these exoplanets at 500 nm. We find that using an optimized high pass filter is an effective way to fit and subtract exozodiacal disk structure while preserving the planetary signal. This method is particularly powerful for low-to-moderately inclined systems with debris disks up to approximately an order of magnitude denser than the habitable zone dust in our Solar System.

In addition to the physical properties of the disk itself, we consider observations with an 8 m and 12 m primary mirror diameter, each with a different coronagraph design, as well as ADI and RDI PSF-subtraction techniques, and two planet detection methods including aperture photometry and PSF matching. Our 8 m architecture is analogous to the $\sim 6$ m inscribed diameter recommended by \citet{Astro2020}. We find that increasing the primary mirror diameter from 8 m to 12 m helps ``resolve out'' the extended source of exozodiacal dust, broadly decreasing the relative time to planetary detection in most cases.  The ADI PSF subtraction technique has a clear advantage over RDI, allowing us to subtract exozodiacal dust for zodi levels greater than $5$ zodis. Finally, we find that using the more advanced PSF matching technique over simple aperture photometry may reduce the required exposure times to detect planets in our synthesized images by up to $\sim 20\%$ and $\sim 50\%$ for the 8 m and 12 m cases, respectively.

The median zodi level of nearby Sun-like stars is $3$ zodis, with a 95\% upper limit of 27 zodis \citep{Ertel2020-kt}. Our results suggest that for moderately inclined systems, we may be able to subtract the exozodiacal dust from direct images of these nearby systems to detect Earth-like exoplanets in the habitable zone, even in the wake of worst-case-scenario mean motion resonance structures. However, mitigating exozodi in high-inclination systems remains an open problem.

\section{Acknowledgments}
We thank our anonymous reviewer for their thoughtful and thorough review that improved the clarity and strength of the paper. We acknowledge funding support from the University of Washington's Astrobiology Program, and the Virtual Planetary Laboratory Team, a member of the NASA Nexus for Exoplanet System Science, funded via NASA Astrobiology Program Grant No. 80NSSC18K0829. The simulations in this work were facilitated though the use of advanced computational, storage, and networking infrastructure provided by the Hyak supercomputer system at the University of Washington.

\bibliography{exozodi_bib.bib}

\end{document}